\documentstyle[twocolumn,aps,psfig]{revtex}

\begin{document}
\draft

\twocolumn[\hsize\textwidth\columnwidth\hsize\csname @twocolumnfalse\endcsname

\title{Magnetization plateau in $S=3/2$ antiferromagnetic 
Heisenberg chain\\
with anisotropy}

\author{T\^oru Sakai$^{1}$ and Minoru Takahashi$^{2}$}
\address{
$^{1}$Faculty of Science, Himeji Institute of Technology, Kamigori,
Ako-gun, Hyogo 678-12, Japan\\
$^{2}$Institute for Solid State Physics, University of Tokyo, Roppongi,
Minato-ku, Tokyo 106, Japan\\
}

\date{October 97}
\maketitle 

\begin{abstract}
The magnetization process of the $S$=3/2 antiferromagnetic Heisenberg chain 
with the single-ion anisotropy $D$ at $T=0$ is investigated by the exact
diagonalization of finite clusters and finite-size scaling analyses. 
It is found that a magnetization plateau appears at $m=1/2$ for 
$D>D_c=0.93 \pm 0.01$. 
The phase transition with respect to $D$ 
at $D_c$ is revealed to be 
the Kosterlitz-Thouless-type. 
The magnetization curve of the infinite system is also presented 
for some values of $D$.
\end{abstract}

\vskip2pc]
\narrowtext

%
%

One-dimensional antiferromagnets have various quantum effects 
observed even in macroscopic measurements.
The Haldane gap\cite{haldane}, which is the lowest excitation gap of the 1D 
Heisenberg antiferromagnets with integer $S$, 
was also detected as a transition from a non-magnetic state to magnetic
one in high-field magnetization measurements of 
Ni(C$_2$H$_8$N$_2$)$_2$NO$_2$(ClO$_4$), abbreviated NENP, 
which is an $S=1$ quasi-1D antiferromagnet.\cite{nenp1,nenp2}  
Recently Oshikawa,Yamanaka and Affleck \cite{oshikawa} suggested that 
even for the 1D $S=3/2$ (half-odd integer) antiferromagnet 
an energy gap is possibly induced by a magnetic field 
and a magnetization plateau appears at $m=1/2$, 
which corresponds to 1/3 of the saturation moment. 
Their argument is based on 
the analogy to the quantum Hall effect 
and the valence bond solid picture for $S=1$.\cite{vbs}  
The magnetization plateau is also predicted in some 
alternating spin chains\cite{plateau,tonegawa}, 
but the mechanism depends on the structure 
of the unit cell and the argument for them is not 
necessarily valid for uniform chains.  

For the anisotropic $S=3/2$ antiferromagnetic chain, 
a variational approach\cite{tim} gave the phase diagram of the nonmagnetic 
ground state, while few works were done on the magnetic state. 
However, it is easy to understand that it should have 
a magnetization plateau at least 
when the system has the positive and infinitely large 
single-ion anisotropy $D\sum _j (S_j^z)^2$. 
Because in the limit ($D\rightarrow \infty$) 
every site has $S_j^z=1/2$ for the ground state at $m=1/2$ 
and any magnetic excitations changing it into $S_j^z=3/2$ at a site 
have a gap proportional to $D$. 
For finite $D$, however, there is no rigorous proof on the existence of 
the gap at $m=1/2$, 
in contrast to the case of $m\not= 1/2$ in which the system is 
proved to be gapless by the Lieb-Schultz-Mattis theorem.
\cite{oshikawa,lsm} 
Thus some numerical tests are important to check the existence of 
the gap and magnetization plateau at $m=1/2$. 
The density matrix renormalization group approach\cite{oshikawa} 
revealed that 
the isotropic $S=3/2$ antiferromagnetic chain is gapless even 
at $m=1/2$ and a critical value $D_c$ should exist as a boundary 
between the gapless and massive phases. 

In this paper, using the exact diagonalization of finite clusters up to 
the system size $L=14$ and finite-size scaling analyses, 
we investigate the $S=3/2$ antiferromagnetic Heisenberg chain 
with the single-ion anisotropy and 
estimate the critical value $D_c$ at $m=1/2$ 
and determine the universality class of the phase transition 
with respect to $D$.  
In addition we present the ground-state magnetization curve extrapolated
to the thermodynamic limit for 
some typical values of $D$. 

%
%

Consider the 1D $S=3/2$ antiferromagnetic Heisenberg Hamiltonian with 
the single-ion anisotropy in a magnetic field 
\begin{eqnarray}
\label{ham}
&{\cal H}&={\cal H}_0+{\cal H}_Z, \nonumber \\
&{\cal H}_0& = \sum _j {\bf S}_j \cdot {\bf S}_{j+1} 
+D\sum _j(S_j^z)^2, \\
&{\cal H}_Z& =-H\sum _j S_j^z, \nonumber
\end{eqnarray}
under the periodic boundary condition. 
For $L$-site systems, 
the lowest energy of ${\cal H}_0$ in the subspace where 
$\sum _j S_j^z=M$ 
(the macroscopic magnetization is $m=M/L$) is denoted as $E(L,M)$. 
Using Lanczos' algorithm, we calculated $E(L,M)$ 
($M=0,1,2,\cdots,3L/2$) for even-site systems up to $L=14$. 
For finite systems described by the total Hamiltonian ${\cal H}$, 
the energy gap of the magnetic excitation changing the value of $M$ 
by $\pm 1$ is given by 
\begin{eqnarray}
\label{gap}
\Delta _{\pm} \equiv E(L,M\pm 1)-E(L,M) \mp H. 
\end{eqnarray}
If the system is gapless in the thermodynamic limit, 
the conformal field theory(CFT) gives the asymptotic form of the size
dependence of the gap as $\Delta _{\pm} \sim O(1/L)$ 
with fixed $m=M/L$. 
If we define $H_+$ and $H_-$ as 
\begin{eqnarray}
\label{field1}
E(L,M+1)-E(L,M) \rightarrow H_+ \quad (L \rightarrow \infty), \nonumber \\
E(L,M)-E(L,M-1) \rightarrow H_- \quad (L \rightarrow \infty),
\end{eqnarray}
$H_+$ and $H_-$ has the same value
and it gives the magnetic field $H$ for the magnetization $m$ in the 
thermodynamic limit. 
On the other hand, 
if the system has a finite gap even in the limit, 
neither $\Delta _+$ nor $\Delta _-$ vanishes for $L \rightarrow \infty$.
It implies that  
$H_+$ and $H_-$ are different. 
As a result, 
a plateau appears for $H_- < H < H_+$ at $m=M/L$ in the ground-state 
magnetization curve. 

\begin{figure}[htb]
\begin{center}
\mbox{\psfig{figure=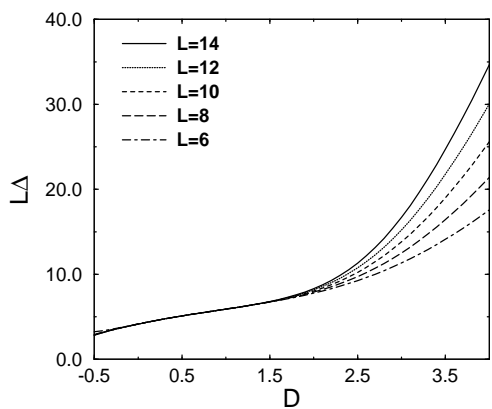,width=7cm,height=6cm,angle=-90}}
\end{center}
\caption{
Scaled gap $L\Delta$ versus the single-ion anisotropy $D$.
\label{fig1}
}
\end{figure}

Since $\Delta _{\pm}$ includes an undecided parameter $H$ 
in the form (\ref{gap}), 
we take the sum $\Delta \equiv \Delta _+ +\Delta _-$ 
for the order parameter of the finite-size scaling, 
to test the existence of the plateau at $m=1/2$. 
(In the massive case, 
the gap $\Delta$ leads to the length of the plateau in the
magnetization curve in the thermodynamic limit.) 
The scaled gap $L\Delta$ of finite systems ($L=6 \sim 14$) at $m=1/2$ 
is plotted versus $D$ in Fig. \ref{fig1}. 
For $D>2$ the scaled gap obviously increases with increasing $L$, 
which means that a finite gap exists in the thermodynamic limit. 
For small $D$ around the region $0<D<1$, 
the scaled gap looks almost independent of $L$. 
It implies that the system is gapless at a finite region. 
At least the form $\Delta \sim 1/L$ is valid for $0\leq D\leq 0.8$ with 
the relative error less than 0.3\% for each point. 
Our precise analysis, however, indicates that 
the $L\Delta$ curves for $L$, and $L+2$ have only one intersection 
in the region $0<D<2$ for each $L$. 
Thus the critical point $D_c$ can be estimated by the phenomenological
renormalization group equation\cite{phenomenological}
\begin{eqnarray}
\label{prg}
(L+2)\Delta_{L+2}(D')=L\Delta_L(D).
\end{eqnarray}
We define $D_{c L,L+2}$ as the $L$-dependent fixed point of (\ref{prg}) and 
it is extrapolated to the thermodynamic limit. 
Fitting the form $D_{c L,L+2} \sim 1/(L+1)$ to the data, 
the extrapolated value is determined as $D_c =0.93 \pm 0.01$,  
based on the standard least square method. 
Thus for $0\leq D< 0.93$ the system is gapless in all the region 
of $0\leq m <3/2$, 
while for $D>0.93$ the energy gap is induced just at $m=1/2$ and 
the magnetization curve has a plateau. 

\begin{figure}[htb]
\begin{center}
\mbox{\psfig{figure=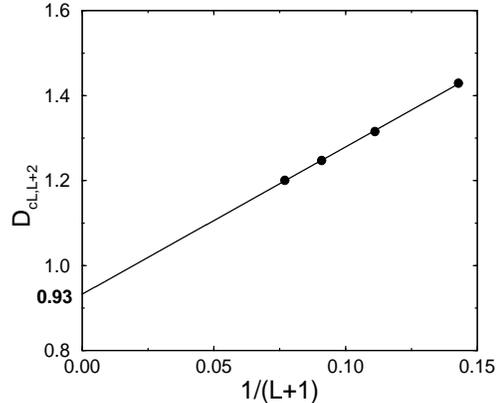,width=7cm,height=6cm,angle=-90}}
\end{center}
\caption{
$L$-dependent fixed point $D_{c L,L+2}$ is plotted versus $1/L$
to determine $D_c$ in the thermodynamic limit.
The estimated value is $D_c=0.93\pm 0.01$.
\label{fig2}
}
\end{figure}

%
%

The phenomenological renormalization group can also estimate the 
exponent $\nu$ defined as $\Delta \sim (D-D_c)^{\nu}$, 
using the $L$-dependent form
\begin{eqnarray}
\label{nu}
\nu_{L,L+2}=\log \big[ {{L+2}\over L} \big] \big/
\log \big[ {{(L+2)\Delta'_{L+2}(D_{c L,L+2})}
\over {L\Delta'_L(D_{c L,L+2})}} \big]
,
\end{eqnarray}
where $\Delta'_L(D)$ is the derivative of $\Delta_L(D)$ with 
respect to $D$. 
The result showed a diverging behavior of $\nu _{L,L+2}$ 
with increasing $L$. 
It implies that $\Delta$ does not have any algebraic form 
near $D_c$. 
Thus the phase transition is expected to be the
Kosterlitz-Thouless(KT)-type\cite{kt}, 
which is also consistent with the existence of 
a finite gapless region under $D_c$. 
In addition 
a naive argument restricting us to three states $S^z=$3/2, 1/2 
and $-1/2$ (neglecting the state $S^z=-3/2$ because of a large 
magnetic field) at each site, leads to a mapping the Hamiltonian
(\ref{ham}) to a generalized anisotropic $S=1$ model without 
magnetic field, which has the KT phase boundary 
between the large-$D$ (singlet) and $XY$ (planar) phases.
\cite{tim,nomura}  

\begin{figure}[htb]
\begin{center}
\mbox{\psfig{figure=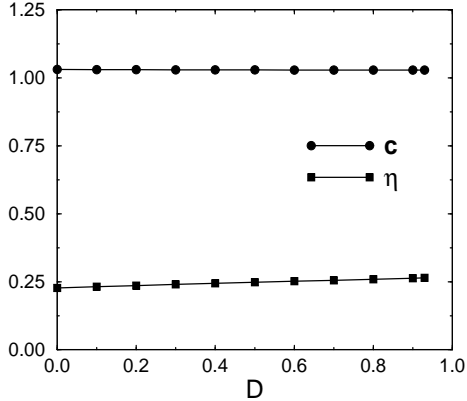,width=7cm,height=6cm,angle=-90}}
\end{center}
\caption{
Estimated central charge $c$ and exponent $\eta$ for
$D\leq D_c$.
At $D=D_c$(=0.93) our estimation gives $c=1.03 \pm 0.06$
and $\eta=0.26 \pm 0.01$.
We conclude $c=1$ for $D\leq D_c$ and $\eta=1/4$ at $D=D_c$.
\label{fig3}
}
\end{figure}

To determine the universality of the phase boundary $D_c$ at $m=1/2$, 
we estimate the central charge $c$ 
in the CFT 
and the critical exponent $\eta$ defined as 
$\langle S_0^+ S_r^- \rangle \sim (-1)^r r^{-\eta}$ 
for $D\leq D_c$. 
The CFT\cite{cft} 
predicts the asymptotic form of 
the ground state energy per site as 
\begin{eqnarray}
\label{gs}
{1\over L}E(L,M) \sim \epsilon (m) -{{\pi} \over 6}cv_s {1\over {L^2}}
\qquad (L \rightarrow \infty),
\end{eqnarray}
where $v_s$ is the sound velocity which is the gradient of the
dispersion curve at the origin. 
Thus the central charge $c$ can be numerically determined by estimating 
the gradient of the plots of $E(L,M)/L$ versus $1/L^2$ and 
$v_s$. 
$v_s$ is estimated by the form\cite{sakai2} 
\begin{eqnarray}
\label{vs}
v_s={L \over {2\pi}}(E_{k_1}(L,M)-E(L,M))+O({1 \over {L^2}}),
\end{eqnarray}
where $k_1=2\pi /L$ is the smallest nonzero wave vector for $L$ 
and $E_{k_1}(L,M)$ is the lowest level in the subspace 
specified by $M$ and $k_1$.  
The calculated $c$ for $D\leq D_c$ at $m=1/2$ is shown 
in Fig. \ref{fig3}. 
At the boundary $D_c(=0.93)$ our estimation gives $c=1.03 \pm 0.06$ 
and other points also have comparable errors. 
Thus we reasonably conclude $c=1$ for $D \leq D_c$. 

Using another prediction of the CFT  
$\Delta _{\pm} \sim \pi v_s \eta/L \quad (L\rightarrow \infty)$, 
the exponent $\eta$ can be estimated by the form\cite{sakai2}
\begin{eqnarray}
\eta ={{E(L,M+1)+E(L,M-1)-2E(L,M)}\over{E_{k_1}(L,M)-E(L,M)}} 
+O({1 \over {L^2}}).
\end{eqnarray}
The calculated $\eta$ is shown in Fig. \ref{fig3}. 
Our estimation $\eta =0.26 \pm 0.01$ at $D=0.93$ suggests 
$\eta =1/4$ just at the phase boundary. 
In addition the estimated $\eta$ gradually decreases with decreasing
$D$.  
Thus the analysis on $\eta$ also supports the KT 
transition. 
 
The critical behavior for $D>D_c$ can be tested by the Roomany-Wyld 
approximation for the Callen-Symanzik $\beta$-function\cite{rw}
\begin{eqnarray}
\label{beta1}
\beta _{L,L+2} (D) = {{ 
1+\log\big({{\Delta_{L+2}(D)}\over {\Delta_L(D)}} \big) \big/
\log \big({{L+2}\over L}\big) }\over{
\big[{{\Delta '_L(D)\Delta '_{L+2}(D)}\over {\Delta_L(D) \Delta_{L+2}
(D)}}\big]^{1\over 2}}}.
\end{eqnarray}
When the gap behaves like $\Delta \sim \exp (-a/(D-D_c)^{\sigma})$, 
the function (\ref{beta1}) has the form 
\begin{eqnarray}
\label{beta2}
\beta _{L,L+2} (D) \sim (D-D_{c L,L+2})^{1+\sigma} 
\quad (L\rightarrow \infty),
\end{eqnarray}
in the thermodynamic limit. 
Fitting the form (\ref{beta2}) to the calculated function
(\ref{beta1}) for each $L$, 
$\sigma$ is estimated as follows: 
$\sigma _{8,10}=0.46 \pm 0.06$, $\sigma _{10,12}=0.52 \pm 0.05$ and 
$\sigma _{12,14}=0.56 \pm 0.06$. 
The results are also consistent with the standard KT
transition ($\sigma =1/2$). 
Therefore we conclude the critical behavior near $D_c$ for $m=1/2$ 
is characterized by the universality class of the 
KT transition. 

%
%
Finally using the method in Ref. \cite{tonegawa,sakai1}, 
we present the ground-state magnetization curve in the thermodynamic
limit for several values of $D$; 
$D=$0, 1, 2 and 3. 
For $D=0$ the system is isotropic and gapless for $0\geq m <3/2$. 
For other cases, it has the gap at $m=1/2$ and the magnetization 
plateau appears.

\begin{figure}[htb]
\begin{center}
\mbox{\psfig{figure=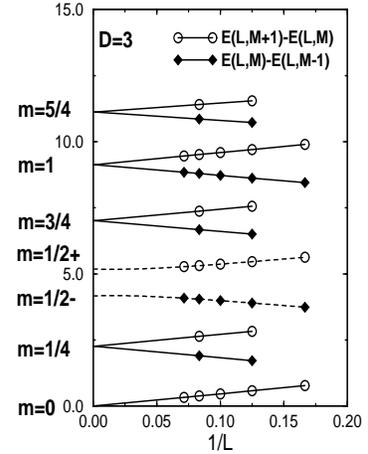,width=7cm,height=7cm,angle=-90}}
\end{center}
\caption{
$E(L,M+1)-E(L,M)$ and $E(L,M)-E(L,M-1)$ plotted
versus $1/L$ with fixed $m$ for $D=3$.
The dashed curves are guides to the eye.
The extrapolated points for $m=1/2-$ and $m=1/2+$ corresponds to
the resutls of the Shanks' transformation $H_-=4.17$ and $H_+=5.19$,
respectively.
\label{fig4}
}
\end{figure}

\begin{figure}[htb]
\begin{center}
\mbox{\psfig{figure=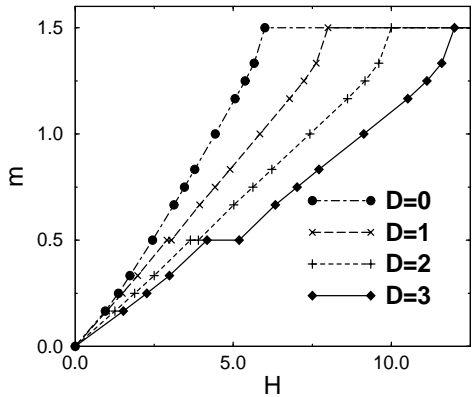,width=7cm,height=6cm,angle=-90}}
\end{center}
\caption{
Ground-state magnetization curves in the thermodynamic limit
for $D=0$, 1, 2 and 3.
\label{fig5}
}
\end{figure}

Since the system is gapless except for $m=1/2$, 
$H_+$ and $H_-$ of (\ref{field1}) 
correspond to each other and the common value 
gives the magnetic field $H$ for given $m$ in the thermodynamic limit.
The size correction of (\ref{field1}) is predicted to decay as $\sim O(1/L)$, 
by the CFT. 
Thus we can estimate $H$ for given $m$, 
using the extrapolation form
\begin{eqnarray}
\label{field2}
E(L,M+1)-E(L,M) \sim H +O(1/L)  \nonumber \\
E(L,M)-E(L,M-1) \sim H +O(1/L) 
\end{eqnarray}
with fixed $m$. 
For $D=3.0$ 
the left hand sides of the form (\ref{field2}) calculated for 
$m=0,1/4,1/2,3/4,1$ and 5/4 are plotted versus $1/L$ in Fig. \ref{fig4}.  
It shows that the form (\ref{field2}) is valid except for $m=1/2$ 
and the two extrapolated values of $H$ 
(the one is extrapolated from $E(L,M+1)-E(L,M)$ and the other is 
from $E(L,M)-E(L,M-1)$) 
correspond to each other well. 
Thus we take the mean value of the two for the magnetic field for each
$m$.  
Only for $m=1/2$ $H_+$ and $H_-$ are obviously different 
and the size correction decays faster than $1/L$,  
as shown in Fig. \ref{fig4}, because the system has a gap. 
Then we estimate $H_+$ and $H_-$ 
by the Shanks' transformation\cite{shanks} 
$P'_n=(P_{n-1}P_{n+1}-P_n^2)/(P_{n-1}+P_{n+1}-2P_n)$ 
for a sequence $\{P_n\}$. 
Applying it twice to $E(L,M+1)-E(L,M)$ and $E(L,M)-E(L,M-1)$ respectively, 
for $L=6,8,10,12$ and 14, 
results in  $H_+=5.19 \pm 0.07$ and $H_-=4.17\pm 0.07$, 
which are indicated as the extrapolated points in Fig. \ref{fig4}. 
The extrapolated value $H$ for other values of $D$  
can be estimated in the same way.  
Only for $D=0$ $H_+$ and $H_-$ correspond even at $m=1/2$. 
The ground-state magnetization curve in the thermodynamic limit 
is given by all the extrapolated values of $H$ for each $m$. 
We present the results for $D=$0, 1, 2 and 3 in Fig. \ref{fig5}, 
where we also used the values of $H$ for $m=$1/3, 2/3, 5/6, 7/6 and 4/3 
which are estimated by the same method as mentioned above. 
The curve has a plateau at $m=1/2$ ($H_- <H<H_+$) for $D=$1, 2 and 3, 
in contrast to the case of $D=0$ which does not have 
any nontrivial behaviors. 

Among those curves in Fig. \ref{fig5}
$D=1$ is the most important in terms of experiments to detect the
plateau,
because $D\sim J$ might be realized in some real materials. 
The candidates of the quasi-1D $S=3/2$ antiferromagnet are 
CsVCl$_3$\cite{vcl} and AgCrP$_2$S$_6$\cite{ag}. 
In particular for AgCrP$_2$S$_6$ a large anisotropic effect 
was observed in the magnetization measurement in low fields. 
Higher-field measurements of those materials would be interesting. 
Note that for $D>D_c$ the ground state is gapless for $H\leq H_-$ and 
$H\geq H_+$, while massive for $H_-<H<_+$. 
In the quasi-1D systems, 
some canted N\'eel orders occur in the 1D gapless phase, 
due to interchain interactions. 
Thus a re-entrant transition might be observed in the magnetization 
measurement; with increasing $H$ the N\'eel order disapears at $H_-$ 
and appears again at $H_+$ at sufficiently low temperatures. 

%
%
In summary 
the finite cluster calculation and size scaling study showed that 
the anisotropic $S=3/2$ has the magnetization plateau 
at $m=1/2$ for $D>D_c=0.93$ and 
the phase transition with respect to $D$ belongs to 
the same universality class as the Kosterlitz-Thouless transition.  

%
%

We wish to thank Dr. M. Yamanaka and Prof. K. Nomura for fruitful 
discussions. 
We also thank 
the Supercomputer Center, Institute for
Solid State Physics, University of Tokyo for the facilities
and the use of the Fujitsu VPP500.
This research was supported in part by Grant-in-Aid 
for the Scientific Research Fund from the Ministry 
of Education, Science, Sports and Culture (08640445).

\end{document}